\documentstyle[preprint,revtex]{aps}

\begin{document}
\draft
\preprint{UBC-TP-92-011; April 9, 1992}

\begin{title}
BOSONIC CHERN--SIMONS FIELD THEORY\\
 OF ANYON SUPERCONDUCTIVITY
\end{title}

\vskip 1in

\author{\bf Nathan Weiss}

\vskip 45pt

\begin{instit}
{Department of Physics, University of British Columbia,\\
Vancouver, B.C. V6T 2A6, Canada}
\end{instit}

\vskip 1.4in

\begin{abstract}
We study the Quantum Field Theory of nonrelativistic
bosons coupled to a Chern--Simons
gauge field at nonzero particle density. This field theory is
relevant to the study of anyon superconductors in which the
anyons are described as {\bf bosons} with a statistical interaction.
We show that it is possible to find a mean field solution
to the equations of motion for this system which has some of
the features of bose condensation. The mean field
solution consists of  a lattice of vortices each carrying a single
quantum of statistical magnetic flux. We speculate on the effects
of the quantum corrections to this mean field solution. We argue
that the mean field solution is only stable under quantum corrections
if the Chern--Simons coefficient $N=2\pi\theta/g^2$ is an integer.
Consequences for anyon superconductivity are presented.
A simple explanation for the Meissner effect in this system is
discussed.
\end{abstract}

\narrowtext
\vfill
\eject

It is now well established that a gas of charged anyons
in two spatial dimensions at $T=0$ is a
superconductor. The most compelling evidence for this is from
the calculation of the current--current correlator in the Chern--Simons
description of the anyon gas. This calculation, which was first
done by Kalmeyer and Laughlin\cite{KalLau} in the RPA approximation
and since worked on by many authors\cite{Lau,Chen,LSW},
finds a massless
pole in the current--current correlator when the anyons are electrically
neutral. This implies that a neutral gas of anyons is a superfluid
and, as a consequence, a charged gas of anyons exhibits a Meissner effect
and is a superconductor.

In the field theoretic work described above one treats the anyons as
fermions coupled to a Chern--Simons (statistical) gauge field. It is this
coupling which gives the fermions anyonic statistics. It is well known
that anyons can be described either as bosons or as fermions\cite{AnyonRev}
 coupled to a Chern--Simons field. The coupling, in each case can
be adjusted to give the desired statistics. Although most of the
field--theoretic work treats the anyons originally as fermions there
are many interesting approaches to anyon superconductivity which begins
with anyons as bosons\cite{WenZee,oth}.
Unfortunately it has not been possible to find an appropriate
field--theoretic analysis to treat a gas of anyons using the
scalar Chern--Simons description.

The main problem blocking such a description is that a noninteracting
system of bosons at nonzero density undergoes bose condensation (at $T=0$ in
$2+1$ dimensions). This makes a mean field (and consequently
an RPA) description of the system coupled to a Chern--Simons
field very difficult. It is the goal of this paper to study this
question in detail and to derive the consequences of a mean field
analysis for anyons which are described as scalars coupled to a Chern--Simons
field at nonzero density. We shall see that no homogeneous
mean field solution exists and that in mean field theory the
system is unstable to the formation of a kind of lattice of
vortices\cite{JK,Olesen}.
We shall present several arguments which suggests that this system
of vortices should be a superconductor. This approach differs
from that of Refs. \cite{WenZee,oth} in that
we seek a mean field solution to the scalar field theory which
can potentially be improved by RPA or other methods.

We thus consider a gas of nonrelativistic anyons with mass $m$
which can be described by a complex bosonic
field $\psi$ coupled to a Chern--Simons
``statistical'' gauge field
$A_\mu$.  We introduce a chemical potential $\mu$ and assume a
Lagrangian of the form
\equation
	i\psi^\dagger\left(\partial_0-igA_0\right)\psi
	-{1\over{2m}}\vert\left(\nabla-ig\vec A\right)\psi\vert^2
	-{{\theta}\over 2}\epsilon_{\mu\nu\lambda}A^\mu\partial^\nu A^\lambda
	-{\lambda\over 2}\left(\psi^\dagger\psi\right)^2
	+\mu\psi^\dagger\psi
	\label{lag1}
\endequation
Notice that besides the interaction introduced by the Chern--Simons
field we have introduced an additional point--like interaction
${\lambda\over 2}\left(\psi^\dagger\psi\right)^2$. The reason for this
is that in the absence of the Chern--Simons coupling, the above system
is known to bose condense. The expectation value of $\psi$ is
$<\psi>^2=\mu/\lambda$. Thus, as is well known, the system is
unstable (unless $\mu=0$ in which case any density is possible).
The quartic term is thus a mechanism for allowing us to use
the chemical potential as a measure of the density.
We shall see however that in the presense of the Chern--Simons
interaction this will not be necessary as the Chern--Simons interaction
itself stablizes the system at nonzero $\mu$. (This is even the
case if the quartic interaction is attractive as has been pointed
out by Jackiw {et. al.}\cite {JK}).

We begin by trying to find a  mean field solution to the equations
of motion derived from Eq. (\ref{lag1})
\equation
	i\left(\partial_0-igA_0\right)\psi=
	-{1\over{2m}}\left(\nabla-ig\vec A\right)^2\psi-\mu\psi
	+\lambda\left(\psi^\dagger\psi\right)\psi
	\label{eom1a}
\endequation
\equation
	g\psi^\dagger\psi=\theta B \equiv \theta \nabla\times A
	\label{eom1b}
\endequation
\equation
	{{ig}\over{2m}}\left(\psi^\dagger D_i\psi-\left(D_i\psi\right)^\dagger
	\psi\right)=\theta\epsilon_{ij}E_j
	\label{eom1c}
\endequation
where $E$ and $B$ are the statistical electric and magnetic fields derived
from $A$ and $D_j=\partial_j-igA_j$ is the covariant derivative.

First note that the only convariantly constant solution to the above
equations (i.e. a solution for which $D_\mu\psi=0$ ) is $\psi=0$.
This is so despite the fact that the chemical potential term drives
$\psi$ away from $\psi=0$. To see this
we decompose $\psi$ into its amplitude $\rho$ and its phase $\Omega$
\equation
	\psi=\rho{\rm e}^{i\Omega}
	\label{phase}
\endequation
$D_\mu\psi=0$ now implies that $\partial_\mu\rho=0$ and that
$gA_\mu=\partial_\mu\Omega$. It follows that $B=\nabla\times A=0$.
But Eq. (\ref{eom1b}) now forces $\rho^2=0$.  It thus follows
that $\psi=0$.  The physical reason for the lack of a covariantly constant
nonzero solution is  that there is an analog of the Meissner effect
at work.  A nonzero $\psi$ which is favored by the chemical potential
term forces (due to the Chern--Simons interaction) a nonzero magnetic
field which is expelled in the presence of a constant nonzero $\psi$ field.

In addition to the lack of solutions
with $D_\mu\psi=0$ there are no covariantly static solutions
with $D_0\psi=0$.
$D_0\psi=0$ now implies that $\partial_0\rho=0$ and that
$gA_0=\partial_0\Omega$. Using this, and the fact that $E_i=\partial_0
A_i-\partial_iA_0$ we find from Eq.(\ref{eom1c}) that
\equation
	E_i=\partial_0\left(A_i-\partial_i\Omega\right)
	\propto \epsilon_{ij}\left(A_i-\partial_i\Omega\right)
	\rho^2
\endequation
Since $\rho$ is time independent it follows that $A-\nabla\Omega$
oscillates in time from which it follows that $B=\nabla\times A$
depends on time and thus from Eq.(\ref{eom1b}) $\rho$ depends on
time. This is a contradiction unless $\psi=0$.

Despite the fact that no covariantly constant solution exists, it is
possible to find a solution for which $\partial_0\psi=0$ with
$A_0\ne 0$. As is shown in Ref. \cite{JK} this can be done by
minimizing the energy functional
\equation
	H=\int d^2x \left[{1\over{2m}}\vert D_i\psi\vert^2
	-\mu\vert\psi\vert^2 +{\lambda\over 2} \vert\psi\vert^4\right]
	\label{energy1}
\endequation
subject to the constraint (Eq.(\ref{eom1b})) that
\equation
	\nabla\times A={g\over\theta}\vert\psi\vert^2
	\label{constraint1}
\endequation
The term proportional to $A_0$ in Eq.(\ref{eom1a}) appears when
$A_i$ in Eq.(\ref{energy1}) is varied with respect to $\psi$ using
the constraint (\ref{constraint1}). This occurs
{\bf provided} $A_0$ is given by
the solution to Eq.(\ref{eom1c}) with $\partial_0 A_i=0$.
It is now clear that a spatially constant nonzero $\psi$ (as opposed to
a {\it covariantly} constant $\psi$ discussed above)
will yield a configuration with infinite energy since the gauge
potential for a constant magnetic field necessarily diverges at
spatial infinity.

There are now two possibilities. Either $\psi=0$ is the mean field
solution of lowest energy or there exists a lower energy non homogeneous
mean field solution. We shall see that at least
for $\lambda\ge0$ the latter option holds. In fact we shall explicitly
construct field configurations with energy $H<0$.

In general it is not possible to actually find the minimum of
the energy for this highly nonlinear system. One approach
is to tune $\lambda$ so that there exist
solutions of the Bogomolny\cite {Bogomol} type to this system. This requires
$\lambda < 0$ i.e. a repulsive interaction. Solutions of
this type have been studied in detail by Jackiw {et. al.}\cite{JK}
and by Olesen\cite{Olesen} in the absence of the chemical potential
term. They do find exact solutions whose structure is that
of a lattice of vortices. In this letter we discuss the
more general case when an exact Bogomolny solution is not available.
The system will be studied using variational methods.

Our first example of a configuration with lower energy than
that of the $\psi=0$ solution consists of a single ``vortex'' with
a total flux $\Phi$.
Consider, as a variational ansatz, a configuration for which the radial
component of the gauge potential $A_r=0$ and whose azimuthal component
\equation
	A_\phi={\Phi\over{2\pi g}}~{{h(r)}\over {r}}
\endequation
so that
\equation
	B={g\over\theta}\rho^2={\Phi\over {2\pi gr}}{{dh(r)}\over {dr}}
\endequation
Since $\rho^2>0$ it follows that $B$ is always positive and that
$h(r)$ is a monotonically increasing function of $r$. We further
demand that $h(r)$ vanish as $r\rightarrow 0$ so that there is no
additional delta function source of magnetic flux at the origin.
Since the total flux is simply $g\int B~d^2r=\Phi\times h(\infty)$
we choose $h(\infty)=1$.

As a simple example choose
\equation
	h(r)=1-{\rm e}^{-(\xi r)^\beta}\left(1+(\xi r)^\beta\right)
	\label{hexample}
\endequation
so that $h(r)\sim (\xi r)^{2\beta}/2$ as $r\rightarrow 0$.
For a single vortex there is no need to choose $\Phi$ to be
a multiple of $2\pi$. We shall see later that in the case
of a lattice of vortices, this will be essential. In anticipation
of this fact we shall consider here a vortex whose flux is
$2\pi$. In this case we see that for large $r$
\equation
	gA_\phi\rightarrow {1\over r}~~~{\rm as}
	~r \rightarrow \infty
\endequation
We can eliminate the effects of this $1/r$ tail by letting the
phase $\Omega$ of the field $\psi$ (Eq.({\ref{phase}})) be given by
\equation
	\Omega=\phi
\endequation
where $\phi$ is the azimuthal angle. This then implies that
$\nabla \Omega = \hat\phi/r$ so that $g\vec A -\nabla \Omega$ falls
off exponentially for large $r$. This will turn out to be essential
for the multi vortex solutions.

Using this ansatz we can now compute the energy
of this configuration. For simplicity we consider the case for which
$\lambda=0$. We shall discuss the more general case in a future
publication.  Using Eq.(\ref{energy1}) the energy
can be written as
\equation
	H=\int d^2x \left\{ {1\over{2m}} \left[\left( \partial_i\rho\right)^2
	+\rho^2\left(gA_i-\partial_i\Omega\right)^2\right]
	-\mu\rho^2\right\}
	\label{energy2}
\endequation
This integral can be done numerically for arbitrary $\beta$ and the
value of $\beta$ which minimizes the energy for fixed $\xi$ can be
found. This value turns out to be $\beta\sim 1.7$.
(One should not take this value of $\beta$ too seriously since
it applies to the single vortex case only and not to the case
of a lattice of vortices which is of greater interest.) At this value
of $\beta$ the expression for $H$ is given by
\equation
	H\sim N\left\{ {\xi^2\over{2m}}\times 1.5 -\mu\right\}
\endequation
where the factor $1.5$ is the approximate numerically obtained value.
Note that for sufficiently small $\xi$ the energy is lowered below
$H=0$ which is the value of $H$ for $\psi=0$. We thus see that there
exist lower energy  solutions than the $\psi=0$ solution.
The energy is lowest when $\xi \rightarrow 0$ which corresponds
to a very large (or diffuse) flux tube. In the case of a lattice
of vortices we shall see that the tubes do prefer to have a definite
finite size.

It is interesting to note that these vortices differ significantly
from the usual Abrikosov vortices found in superconductors.
First of all the order parameter $\rho$ vanishes outside the
vortex. Secondly the magnetic field tends to zero as
$r^{2\beta-2}$ at the centre of the vortex. Both of these
peculiar properties are a result of the Chern--Simons condition
which forces $B$ to be proportional to $\rho^2$.

We now move on to multivortex solutions which we shall show lower
the energy even more.  In fact for a uniform distribution of
vortices we shall see that the mean energy density is less than
zero. Our technique, as for the single vortex, will be to
construct a variational ansatz for a multivortex configuration.
We imagine a  distribution of $n$ vortices at locations
$\vec r_1 \cdot\cdot\cdot \vec r_n$. Recall from the single vortex
solution that it is imperative that $\psi^\dagger\psi$ vanish at
the location of each vortex so that the integral of
$(\partial_i\Omega)^2\rho^2$ converges there. Furthermore
no vortex can have a long range $1/r$ component which is uncancelled
by a $\partial_i\Omega$ term. If it did, then the integral of
$(gA_i-\partial\Omega)^2\rho^2$  over all the other vortices
would diverge. The only option is for each vortex to have a flux
of $2\pi\times$ an integer so that the $1/r$ piece can be cancelled
at infinity. In general vortices with flux $2\pi$ (i.e. a single
quantum of flux) will provide configurations with the lowest
energy (although we shall see cases where this may not be so).

A multivortex ansatz which satisfies all our requirements is
\equation
	g\vec A(\vec r)=\left(\sum_{i=1}^N{{\hat\phi_i}\over
	{\vert \vec r-\vec r_i\vert}}\right)\times
	\prod_{j=1}^N h\left(\vert \vec r-\vec r_j\vert\right)
\endequation
Since $h\rightarrow 1$ for large $r$ we see that this ansatz behaves
like a single vortex near each of the vortices.
To cancel the long range behavior of $\vec A$ we choose the phase $\Omega$
of $\psi$ to be
\equation
	\Omega\left(\vec r\right)=\sum_{i=1}^N \phi\left(\vec r,\vec r_i
	\right)
\endequation
where $\phi(\vec r,\vec r_i)$ is the azimuthal angle which the vector
$\vec r-\vec r_i$ makes with some arbitrary axis.
As a result the important combination $gA-\nabla\Omega$ goes like
\equation
	g\vec A -\nabla\Omega = \left(\sum_{i=1}^N{{\hat\phi_i}\over
	{\vert \vec r-\vec r_i\vert}}\right)\left(\prod_{j=1}^N
	h(\left(\vert \vec r -\vec r_j \vert \right)-1 \right)
\endequation
It is now clear that the contribution of this term due to any given
vortex `C' at the location of any other vortex `D' vanishes exponentially
as the distance between `C' and `D' increases.

We could now proceed by choosing a specific form for $h(r)$ as
might be given, for example, by Eq.(\ref{hexample}) and minimizing the
energy functional from Eq.(\ref{energy2}). We can learn much more,
however, by doing a general analysis of the form of the answer.
Let us, for definiteness, imagine that the vortices are arranged
on a square lattice. Let $d$ be the separation between the vortices
and let $1/\xi$ be a measure of the size of the vortex (as is given,
for example, in Eq.(\ref{hexample})). The energy in Eq.(\ref{energy2})
has two pieces. The first piece proportional to $1/m$ and the second
proportional to $\mu$.

Using the fact that $g\rho^2=\theta B$ and the definition of
$N=2\pi\theta/g^2$ and recalling that the total flux per vortex
is $2\pi$ we can compute the second term in Eq.(\ref{energy2}) as
\equation
	\int d^2x~\mu\rho^2=Nn\mu
\endequation
where $n$ is the total number of vortices
\equation
	n={V\over {d^2}}
\endequation
where $V$ is the volume of the system. Evaluation of the first term
in Eq.(\ref{energy2}) is somewhat more involved. The terms proportional
to $1/m$ are functions of only $\xi$ and $d$ and have the dimensions
$[{\rm mass}]^2$. Thus this first term can be written as
\equation
	{1\over {2m}}Nn\xi^2\times F(\xi d)
\endequation
where the function $F$ is unknown until the integrals have been done
explicitly. But nonetheless it is a function of the dimensionless product of
$\xi$ and $d$\cite{foot}.   Combining these two terms we can write the
total energy
\equation
	H=nN\left({{(\xi d)^2F(\xi d)}\over{2md^2}}-\mu\right)
\endequation

The behavior of $x^2F(x)$ for small and large $x$ is easy to determine.
If $\xi \rightarrow\infty$ the vortices are very small relative to
their spacing and we obtain simply $n$ copies of the single vortex
solution.  As a result $F\rightarrow {\rm constant}$ and thus
$x^2F(x)\rightarrow \infty$ as $x\rightarrow \infty$.
On the other hand when $\xi\rightarrow 0$, $F$ diverges very
severely since the magnetic field is now nearly constant
and the energy thus grows like $V^2$. Thus not only $F$ but also
$x^2F(x)$ diverges as $x\rightarrow 0$.  As a result, $x^2F(x)$
reaches some minimum value as $x$ is varied.  Let $\eta$ be this
minimum value of $x^2F(x)$.  It is clear that $\eta$ will be
a number of order 1. The precise value depends on the precise form of
$h(r)$ and it can be computed.
It follows that for this minimum value of $\xi d$
\equation
	H_{\rm min}={{VN}\over {d^2}}
        \left\{ {{\eta}\over{2md^2}}-\mu\right\}
\endequation
Note that $d$ is still a variational parameter. The minimum value
of $H_{\rm min}$ occurs when $d^2=\eta/m\mu$ in which case
\equation
	H_{\rm min}= -V{{m\mu^2N}\over{2\eta}}
\endequation
Note that the condensate provides us with a negative energy density
which is thus lower than the energy for the $\psi=0$ solution.

We conclude that in the context of the variational calculation
it is evident that a ``condensate'' of vortices provides a good
candidate for the Mean Field ground state of this system.

But this is not the whole story! Let us now compute the charge
$Q$ of one of these vortices in the ``lattice''.
\equation
	Q=g\int d^2x~\psi^\dagger\psi=Ng
\endequation
There are $N$ anyons in this vortex with a total charge $Ng$.
This is fine if $N$ is an integer. But if $N$ is not an integer
then $Q$ is not a multiple of $g$ {\bf and} the number of particles
in the vortex is not an integer. Although the mean field solution
still exists in this case we should expect to run into serious trouble when
we consider quantum fluctuations about this vortex configuration.
To see this let us transport one vortex around another vortex.
The Bohm--Aharonov phase picked up is now
\equation
	{\rm e}^{igN\int Bd^2x}={\rm e}^{2\pi iN}\ne 1
	~~~~~{\rm if}~{\rm N~is~not~an~integer}
\endequation
Thus if $N$ is not an integer these vortices are themselves anyons.
If $N$ is rational ($N=p/q$ with $p,q$ integers) then it may be
possible to consider a mean field solution with flux $2\pi q$ per
vortex. But if $N$ is irrational then it seems clear that the
quantum corrections will lead to infinities.
It is also likely that quantum corrections will force the number
of particles in the vortex to be an integer. They may also
 ruin the lattice structure of the mean field
solution and cause the system to form a liquid.

Let us now concentrate on the case in which $N$ is an integer.
We must consider separately the case when $N$ even and when
$N$ is odd. In the case $N$ even (which includes the case of semions
for which $N=2$) the vortices are bosons. When one vortex is taken
around another, the phase is ${\rm exp}(4\pi i)$ which leads
to a phase ${\rm exp}(2\pi i)$ under interchange. These bosons
have a repulsive core. It is well known that bosons with a repulsive
core in two space dimensions form a superfluid.
(At nonzero temperature there is no long range order in 2+1 dimensions
but the superfluidity persists up  to some critical
(Kosterlitz--Thouless) transition temperature.) If $N$ is odd the
vortices are fermions. It is likely that when quantum effects
are included the energy will be lowered by combining two vortices
into a single vortex with flux $4\pi$. This new mean field vortex
will then be a boson which will allow for bose condensation and
superfluidity.

It is possible to see from the mean field theory directly how the
Meissner effect arises in this theory. To this end let us couple
our system to a ``real'' fixed background
electromagnetic field ${\cal A}_\mu$.
If ${\cal A}_0=0$ then the mean field problem can be formulated
as a search for the minimum of
\equation
	H=\int d^2x\left[{1\over{2m}}\left\vert\left(\partial_i
	-igA_i-ie{\cal A}_i\right)\psi\right\vert^2-\mu\vert\psi
	\vert^2\right]
	\label{energy3}
\endequation
The mean field solution which we found for $e=0$ will certainly not
admit a constant magnetic field $b=\nabla\times{\cal A}$.
If we choose, for example, a vector potential ${\cal A}_2=bx_1$ with
${\cal A}_1=0$ then Eq.(\ref{energy3}) will have a term
$\int d^2x \rho^2(\vec r)x_1^2$ which clearly diverges due to the
presence of the infinite lattice of vortices.

The above argument is however too crude since it is possible
to make new kinds of vortices in the presence of an externally
applied magnetic field. One can simply force the {\bf sum} of the
flux of the statistical and the real gauge fields to be $2\pi$ so that
\equation
	\int d^2x\left[g\nabla\times A +e\nabla\times{\cal A}\right]=2\pi
\endequation
while $\rho^2$ is proportional to only the statistical magnetic
field $B=\nabla\times A$. This will yield a perfectly sensible mean
field solution for arbitrarily small applied magnetic field. The
problem with this solution is that the charge of the vortex will no
longer be an integral multiple of $g$
\equation
	Q=\int d^2x \rho^2={{gN}\over{2\pi}}\int d^2x\nabla\times A
	\ne gN
\endequation
and, more importantly, the total number of particles in each
vortex will no longer be an integer.
As discussed previously the quantum corrections should cause such
a configuration to have an infinite energy.
As a result, a constant magnetic field is expelled from this system.

Although the bosonic picture of anyonic superconductivity
looks on the surface to be dramatically different from the
fermionic description there are many similarities. In fact many
of the predictions of this model are very similar to those
for the fermionic description. For example a crude calculation of
the real magnetic field generated by these anyons which should
be observable in muon spin rotation experiments
agrees quite well with the result obtained in the fermion analysis.
Details of this calculation will be given in a future publication.
The one major difference between this analysis and the fermionic
RPA analysis is in the behavior of the system at nonzero temperature.
In the fermionic analysis the RPA calculation alone leads to
a loss of superconductivity at nonzero temperature\cite{LSW}.
In the bosonic case we have a conventional Kosterlitz--Thouless
kind of superconductor which continues to be a superconductor
up to some critical temperature.

In this paper we have discussed a nonrelativistic theory of
Bosons coupled to a Chern--Simons gauge field.
It is also possible to carry out the analysis of a {\bf relativistic}
Chern--Simons model at nonzero density. The details of that
analysis differ considerably from the nonrelativistic analysis
but the conclusions remain the same. The system forms a
lattice of vortices which expected to be a superfluid. Details
of the relativistic analysis will also be presented in a future publication.

\vskip .3in
\centerline{\bf Acknowledgements}

This work was supported in part by the Natural Sciences and Engineering Council
of
Canada.  I wish to thank Gordon Semenoff, Jacob Sonnenschein Joe
Lykken and Alex Kovner for many informative discussions.

\vfil
\eject


\begin{references}
\bibitem{foot} The situation is a bit more subtle than is described
in the text. There is, in fact also a dependence on the volume
$V$ since in the limit where $\xi\rightarrow 0$ the magnetic field
becomes nearly constant and the energy grows like $V^2$. Fortunately
we need not worry about this as long as the minimum of $x^2F(x)$ occurs
for x not too small. This is nearly always the case.
\end{references}
\end{document}